\documentclass[secnumarabic, twocolumn, graphics,floatfix, superscriptaddress,tightenlines, aps, prb]{revtex4-1}

\usepackage[dvips]{graphicx}
\usepackage{dcolumn}
\usepackage{bm}
\usepackage{color}

\begin{document}
\title{Spin temperature concept verified by optical magnetometry of nuclear spins}

%% Notice placement of commas and superscripts and use of &
%% in the author list

\author{M.~Vladimirova}
\affiliation{ Laboratoire Charles Coulomb, UMR 5221 CNRS-Universit\'{e}  de Montpellier,
F-34095, Montpellier, France}

\author{S.~Cronenberger}
\affiliation{ Laboratoire Charles Coulomb, UMR 5221 CNRS-Universit\'{e}  de Montpellier,
F-34095, Montpellier, France}

\author{D.~Scalbert}
\affiliation{ Laboratoire Charles Coulomb, UMR 5221 CNRS-Universit\'{e}  de Montpellier,
F-34095, Montpellier, France}

\author{I.~I.~Ryzhov}
\affiliation{ Spin Optics Laboratory, St. Petersburg State University, 1 Ul'anovskaya,
Peterhof, St. Petersburg 198504, Russia}

\author{V.~S.~Zapasskii}
\affiliation{ Spin Optics Laboratory, St. Petersburg State University, 1 Ul'anovskaya,
Peterhof, St. Petersburg 198504, Russia}

\author{{G.~G.}~Kozlov}
\affiliation{ Laboratoire Charles Coulomb, UMR 5221 CNRS-Universit\'{e}  de Montpellier,
F-34095, Montpellier, France}

\author{A. Lema\^{\i}tre}
\affiliation{ Centre de Nanosciences et de nanotechnologies - CNRS - Universit\'{e} Paris-Saclay - Universit\'{e} Paris-Sud, Route de Nozay, 91460 Marcoussis, France}

\author{K.~V.~Kavokin}
\affiliation{ Spin Optics Laboratory, St. Petersburg State University, 1 Ul'anovskaya,
Peterhof, St. Petersburg 198504, Russia}
\affiliation{ Ioffe Physico-Technical Institute of the RAS, 194021 St.Petersburg, Russia}

\begin{abstract}
%
%The concept of nuclear spin temperature is one of the cornerstones of the nuclear magnetism in solids.%\cite{Goldman,AbragamProctor}. 
%It has made possible realisation of the cryogenic cooling into the microKelvin range
%% \cite{Pickett} 
%and observation of nuclear spin ordering in metals and insulators.
%% \cite{Oja, Abragam}.
%%
%However, proving its validity for semiconductor nano- and microstructures is challenging due to the lack of techniques capable of precise sensing of weak nuclear magnetisation in a small volume.
%
%We develop a method of non-perturbative optical control over adiabatic remagnetisation of the nuclear spin system  in semiconductors and apply it to study nuclear spin thermodynamics in GaAs microcavities.
%
We develop a method of non-perturbative optical control over adiabatic remagnetisation of the nuclear spin system  and apply it to verify the spin temperature concept in GaAs microcavities.
The nuclear spin system is shown to exactly follow the predictions of the spin-temperature theory, despite the quadrupole interaction that was earlier reported to disrupt nuclear spin thermalisation.
% \cite{MaletinskyNatPhys2009}. 
%
These findings open a way to deep cooling of nuclear spins in semiconductor structures, with a prospect of realisation of nuclear spin-ordered states
% \cite{Chapellier,Goldman1974,Merkulov1982,Merkulov98} 
for high fidelity spin-photon interfaces. % \cite{Arnold,Stockill,Sun2016,Gao2012}.
\end{abstract}
\pacs{Valid PACS appear here}% PACS, the Physics and Astronomy

\maketitle
%Then the body of the main text appears after the intro paragraph.
%Figure environments can be left in place in the document.
%\verb|\includegraphics| commands are ignored since Nature wants
%the figures sent as separate files and the captions are
%automatically moved to the end of the document (they are printed
%out with the \verb|\end{document}| command. However, tables must
%be manually moved to the end of the document, after the addendum.
%

The concept of nuclear spin temperature is one of the cornerstones of the nuclear magnetism in solids\cite{Goldman,AbragamProctor}. 
It has made possible realisation of the cryogenic cooling into the microKelvin range
 \cite{Pickett} 
and observation of nuclear spin ordering in metals and insulators \cite{Oja, Abragam}.
Such degree of  control of the nuclear spin system (NSS) in semiconductor heterostructures would allow enhancing the efficiency of spin-based information storage and processing  \cite{Arnold,Stockill,Sun2016,Gao2012}.
However,  proving the validity of the spin temperature concept for semiconductor nano- and microstructures is challenging due to the lack of techniques capable of precise sensing of weak nuclear magnetisation in a small volume.
In addition, recent experiments showed that in quantum dots, where  strong quadrupole-induced local fields have been reported, nuclear spin temperature
failed to establish \cite{MaletinskyNatPhys2009}.
In this context, NSS thermalisation sensing in semiconductor heterostructures is one 
the central issues for both fundamental questions related to the realisation of nuclear spin-ordered states, 
and for  potential applications, such as  high fidelity spin-photon interfaces \cite{Arnold,Stockill,Sun2016,Gao2012}.

\begin{figure*}[!]
\begin{center} 
 {  \includegraphics [width=2\columnwidth] {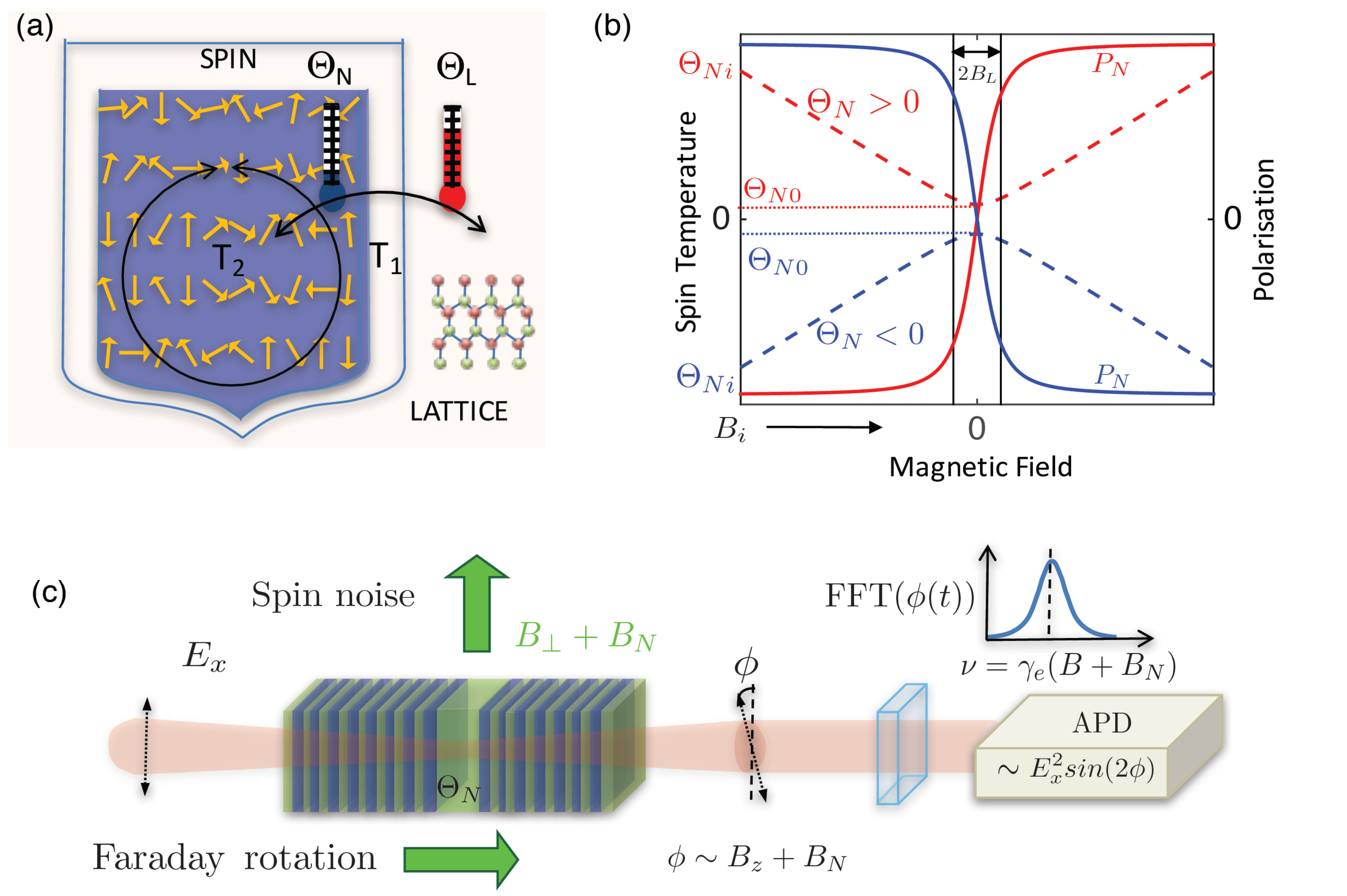} } %\resizebox{5.5in}{!}
 \end{center}
 \caption{ (a)  Sketch of the two heat reservoirs,
  the atomic lattice at temperature $\Theta_L$, and  the nuclear spin system (NSS) at temperature $\Theta_N$. The equilibrium within the NSS is established during the spin-spin relaxation time $T_2<<T_1$, the spin-lattice relaxation time.
%  , so that $\Theta_N \ne \Theta_L$
  (b)  Evolution of the nuclear spin temperature (dashed lines) and polarisation (solid lines) in the adiabatic de(re)-magnetisation process starting from either positive (red lines) or negative initial spin temperature $\Theta_{Ni}$ under magnetic field $B_i$, as described by equation (1).  The lowest nuclear spin temperature $\Theta_{N0}$ that can be reached in the adiabatic demagnetisation procedure
is determined by the initial temperature of the nuclei $\Theta_{Ni}$ in the strong magnetic field $B_i$ and the
local field $B_L$. (c) Schematic view of the sample and the detection stage of Faraday rotation and spin noise experiments. NSS in the cavity probed using two optical technics, that allow us to trace the evolution of the initially prepared nuclear spin polarization $P_N$ and temperature $\Theta_N$ along the demagnetisation process. Spin noise spectrum is obtained as Fourier transformation of the stochastic Faraday rotation. The spectral peak frequency
is directly related to the Overhauser field acting on electrons in the presence of the in-plane magnetic field. }
 \label{fig:sceme}
 \end{figure*}

The basic postulates of the spin temperature theory are illustrated in Fig. \ref{fig:sceme}(a). It is assumed that during the characteristic time $T_2$ determined by spin-spin interactions the NSS reaches the internal equilibrium. This means that properties of the NSS are governed by a single parameter, the spin temperature $\Theta_{N}$. When this temperature is made different from the lattice
temperature  $\Theta_L$ (e.g. by the optical pumping), the thermalisation of the NSS with the crystal lattice usually requires a much longer characteristic time $T_1$.
Fig.1(b) illustrates one of the main predictions of the spin temperature theory: if the NSS is subjected to a slowly varying magnetic field, such that $dB/dt<B_L/T_2$, then $\Theta_N$ and the nuclear spin polarisation $P_N$ change obeying universal expressions:
\begin{equation}
\Theta_N/\sqrt{B^2+B_L^2}=\Theta_{Ni}/B_i;
\quad P_N=\frac{B}{3k_B\Theta_N}\hbar \langle \gamma_N (I+1)\rangle.
\label{eq1}
\end{equation}
%\begin{equation}
%p_N\Theta_N/B=const
%\end{equation}
%
Here $\gamma_N$ is  the gyromagnetic ratio of the nuclear spin $I$, angular brackets denote the averaging over all nuclear species, $k_B$ is the Boltzman constant, and $\Theta_{Ni}$ is the spin temperature at strong magnetic field $B_i>>B_L$, where  $B_L$ is the local field induced by the  fluctuating  nuclear spins.
These generic relations are based on the principle of entropy conservation in a thermodynamic system
during adiabatic process.
They constitute the basis for the nuclear spin cooling by adiabatic demagnetisation,
a widely used  cryogenic technique \cite {Tuoriniemi,Pobell,Kurti,Oja}.
The nuclear spin temperature may take either positive or negative values,
in the latter case  the magnetisation being anti-parallel to the applied field.
%
%The lowest nuclear spin temperature $\Theta_{N0}$ that can be reached in the adiabatic demagnetisation procedure
%is determined by the initial temperature of the nuclei $\Theta_{Ni}$ in the strong magnetic field $B_i$ and the
%local field $B_L$.
%

Various optical and magnetic techniques have been employed to measure nuclear spin temperature, mostly by the magnetisation measurement  at a fixed value of the external magnetic field \cite{Chapellier,Goldman1974,Chekhovich2017,Tuoriniemi,Oja}.
On the other hand, a  direct measurement of the nuclear magnetisation  as a function of slowly
 varying magnetic field  is extremely challenging and has never been realised to the best of our knowledge.
Such an experiment is required to check rigorously the validity of the concept of spin temperature as applied to a specific system.

In this Letter we report on realisation of such a proof-of-concept experiment in microcavities, semiconductor microstructures with enhanced light-matter coupling \cite {kavokin2007microcavities}.
%
%
%\begin{figure}
% \begin{center} { \includegraphics [width=1\columnwidth] {Setup.pdf} } %\resizebox{3.5in}{!}
% \end{center}
% \caption{ {\bf Schematic view of the sample and the detection stage of Faraday rotation and spin noise experiments. \textbar }  NSS in the cavity probed using two optical technics, that allow us to trace the evolution of the initially prepared nuclear spin polarization $P_N$ and temperature $\Theta_N$ along the demagnetisation process.
% %
%Faraday rotation angle of the probe beam in the presence of the longitudinal magnetic field is sensitive to the longitudinal component of the nuclear magnetisation.
%%
%Spin noise spectrum is obtained as Fourier transformation of the stochastic Faraday rotation. The spectral peak frequency
%is directly related to the Overhauser field acting on electrons in the presence of the in-plane magnetic field. }
% \label{fig:setup}
% \end{figure}
%%
The principle of our experiment is sketched in Fig. \ref{fig:sceme}(c). Prior to the measurement, the NSS of the n-GaAs layer embedded in a microcavity
%(quality factor $Q\sim20000$, see Methods)
%
 is polarised by optical pumping in the presence of the longitudinal magnetic field.
Nuclear spin polarisation is probed by linearly polarized cavity mode photons with the photon energy in the transparency band of GaAs.
% ($h\nu \sim E_g-20$~meV, where $E_g=1.51eV$ is the GaAs bandgap).
%
Polarisation of the light beam transmitted through the cavity is sensitive to the {\it Overhauser field}, an effective magnetic field created by NSS and acting on electron spins \cite{OpticalOrientation}.
Two methods of detection of nuclear spin polarisation are used: (i) the Faraday effect induced by the Overhauser field \cite{Artemova1985,Giri2013} and (ii) the  spin noise spectroscopy of resident electrons subject to the Overhauser field \cite{Ryzhov2015,Berski2015,Ryzhov2016}.
\begin{figure*}
 \begin{center}{!}{ \includegraphics  [width=2\columnwidth]  {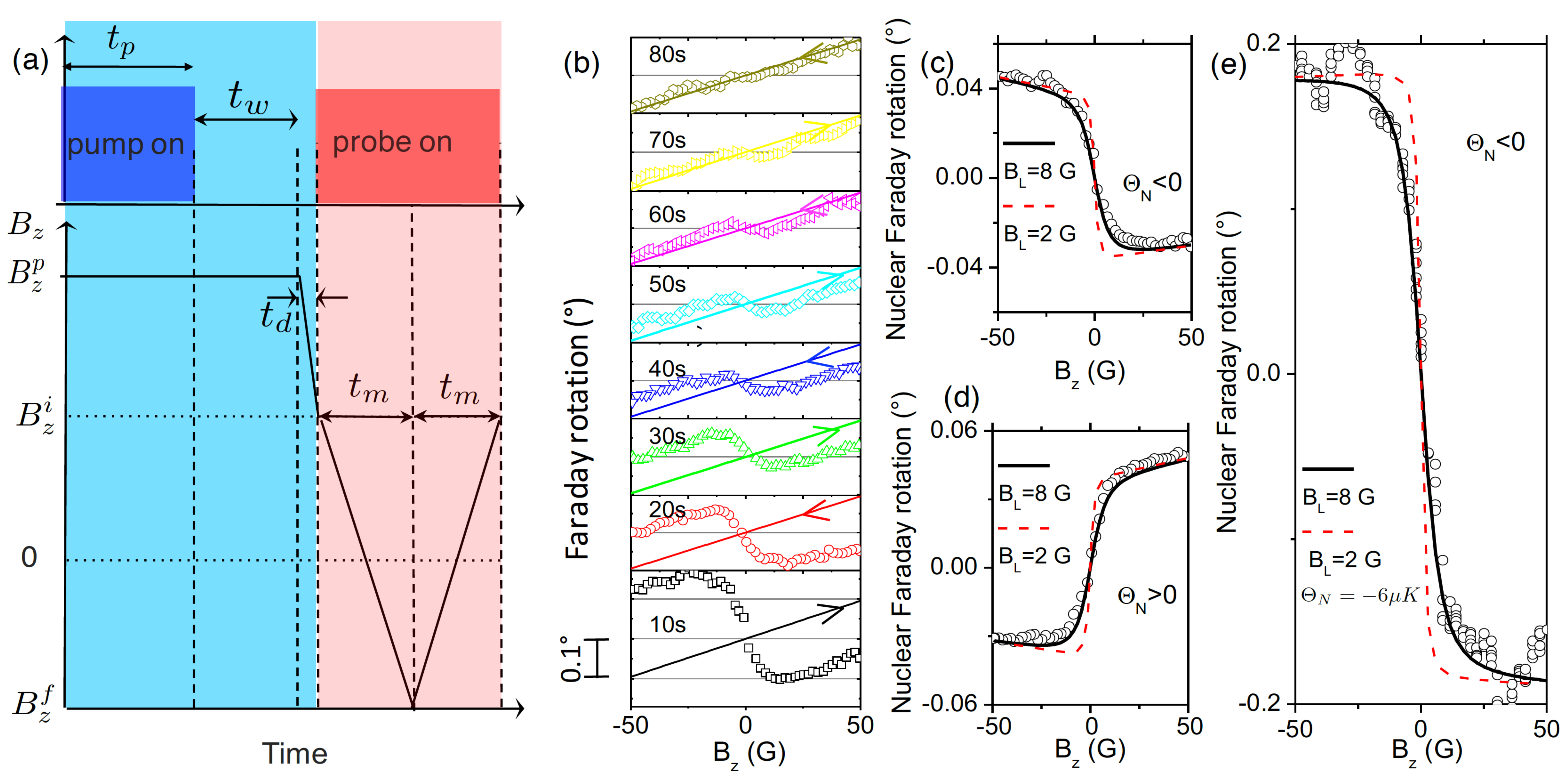} } %%  \resizebox{6.5in}
 \end{center}
 \caption{ Nuclear spin magnetometry by Faraday rotation. (a)
 Timeline of the experiment. The preparation (blue area) consists in pumping under longitudinal magnetic field $B_z^p$, waiting for eventual nuclear relaxation in the vicinity of the localised electrons during $t_w$ and  fast demagnetisation down to $B_z^i$. Faraday rotation of the probe beam is measured during successive scans of the magnetic field across zero (pink area, only first scan is shown).
 (b) Raw measurements of the Faraday rotation in Sample B (circles). NSS is prepared at $\Theta_N<0$. During nine successive scans of the magnetic field ($t_m=5$~s, direction shown by arrows) conventional Faraday rotation remains constant, this contribution is shown by solid lines. The remaining contribution to the signal is due to the nuclear spin polarisation. It is shown separately in   (e) for the first scan.
 (c-d)  Faraday rotation induced in Sample A by nuclear spin prepared either at negative (c) or
 at positive  (d) temperature (circles). Lines in (c-e) are calculated from Eq. (\ref{eq1}), assuming different values of the local field, see Supplemental Material (SM). }
 \label{fig:fr}
 \end{figure*}
The main  features of the behaviour of the optically cooled NSS under varying external magnetic fields are demonstrated in the experiment where the Faraday rotation angle is measured while ramping the longitudinal magnetic field across zero (Fig. \ref{fig:fr}).
The experiment is conducted in two steps: preparation
%, including optical pumping of nuclei at $B_z^p$ and  initial demagnetisation to $B_i$,
and measurement (Fig. \ref{fig:fr} (a)).
% under varing magnetic field $B_z$
%varying between  $B_z^i$ and  $-B_z^i$ .
%
%%%The duration of each scan is  $\sim 10$~s so that $dB/dt=10$~G/s.
%
The  measured signal (Fig. \ref{fig:fr}(b))
%measured during the second step is shown in Fig. \ref{fig:fr}b. It
contains two contributions:
Faraday rotation directly induced by the external field (shown by solid lines, it remains unchanged
for all the scans), and the Faraday rotation induced by the Overhauser field $\phi_N$ ( shown separately in Fig. \ref{fig:fr}(e) for the first scan),
which is proportional to the nuclear spin polarisation.
%%%:
%%% \begin{equation}
%%%\phi_N=B_N V_N L=b_N P_N V_N L,
%%%\label{phiN1}
%%% \end{equation}
%%%where $b_N=5.3$~T is the Overhauser field produced by the fully polarised nuclear spins, $V_N$ is the nuclear Verdet constant, $L$ is the effective optical length of the sample accounting for
%%%by multiple round trips of  light in the cavity.
%
In each consecutive scan, $\phi_N$  diminishes due to the nuclear spin-lattice relaxation, but  the behaviour of nuclear polarisation is described  by Eqs. (1):  the polarisation is an odd function of the applied field, there is no remanent magnetisation at $B=0$,
and $B_L=8\pm2$~G.
  We have performed this analysis for two samples with different concentrations of Si donors $n_d$: an insulating sample  with $n_d=2\cdot10^{15}$~cm$^{-3}$ (Sample A) and a sample characterised by a metallic conductivity ($n_d=2\cdot10^{16}$~cm$^{-3}$, Sample B), for NSS prepared
  either at positive, or at negative spin temperature.
 The value of $B_L$   obtained for  both samples is the same within our experimental accuracy.

\begin{figure*}
 \begin{center} \resizebox{6.5in}{!}{ \includegraphics{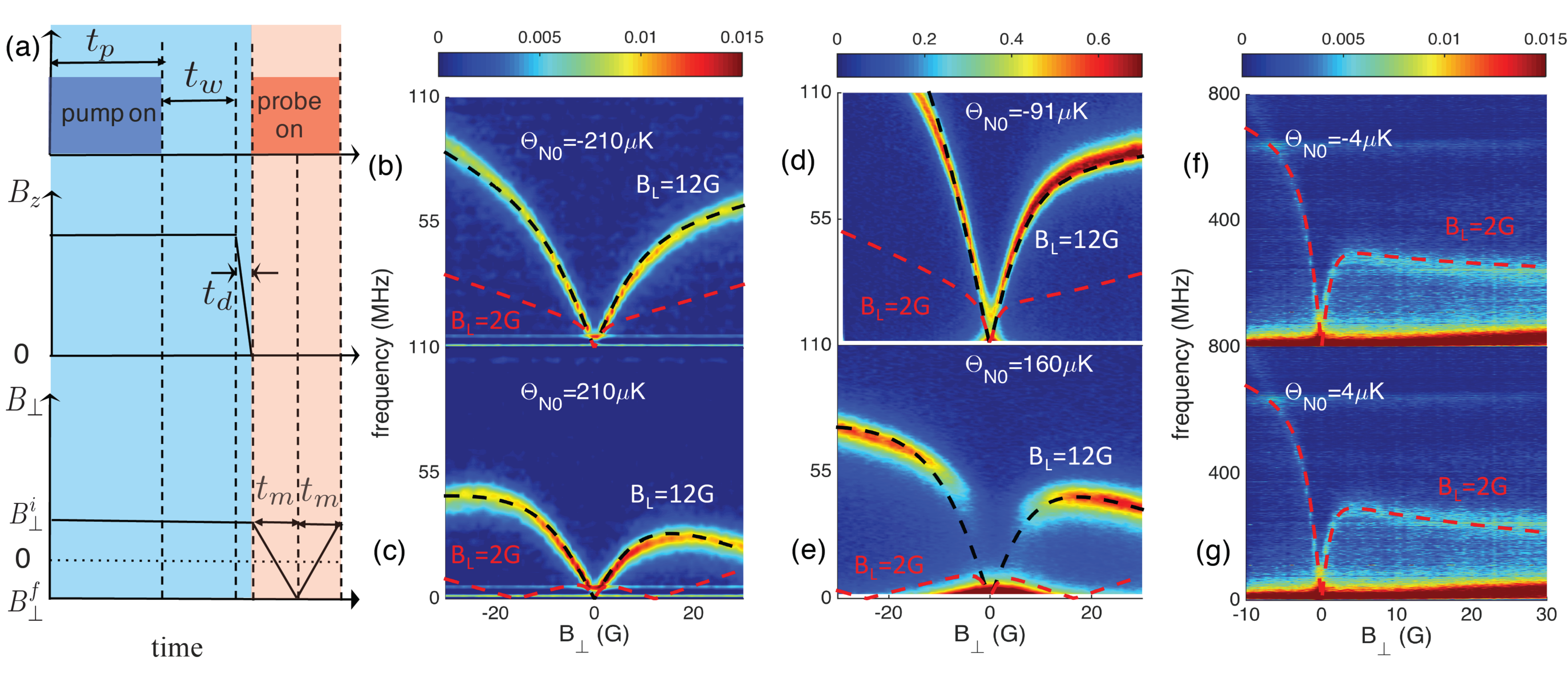} }
 \end{center}
 \caption{ Nuclear spin magnetometry by spin noise spectroscopy.   (a)
 Timeline of the experiment. The preparation (blue area) consists in pumping under oblique magnetic field, waiting during the time $t_w$ required for nuclear relaxation in the vicinity of localised electrons and  fast demagnetisation down to $B_\perp^i$. Spin noise spectra of the probe beam are measured while scanning $B_\perp$ across zero (pink area).
  (b-g) Color maps of the spin noise spectra during adiabatic demagnetisation procedure at positive (c, e and g) and negative (b, d and f) spin temperature (measured in the signal to shot noise ratio units) for two microcavity samples A ( b-c), B (d-e) and a bulk sample C (f-g). Black lines in (b-e) and red line in (f-g)  are fits to Eqs. \ref{eq1}, that determine the values of  $B_L$ and $\Theta_{N0}$ indicated on the figure (see also SM).
 Red lines in (b-e) illustrate how the the value $B_L=2$~G fails to describe the experiment.
   }
 \label{fig:sn}
 \end{figure*}

We  complemented these results by spin-noise measurements of nuclear remagnetisation under magnetic field perpendicular to the light and the structure axis (Fig. \ref{fig:sn}).
% and get an easier access to the values of the spin temperature
%  we have measured the nuclear polarisation under slowly varying magnetic field by the
 %
%In these experiments, the pumping stage is almost identical to the Faraday rotation experiment described above, except for the presence of a small transverse component of the magnetic field $B_{\perp}^i$ (Fig. \ref{fig:sn}a).
% %
%For the measurements of the spin noise spectra, $B_z$ is reduced to zero, and the transverse component of the field is tuned across zero at the rate of $\sim 0.5$~G/s.
 %
Color maps in Figs. \ref{fig:sn}b,c show the evolution of the electron spin noise spectra under varying magnetic fields.
%
%For both Samples A and B, we study NSS prepared at positive and negative temperature.
%
The narrow peak in the spectra appears at the frequency $\nu$ of the electron Larmor precession in the
total effective magnetic field acting upon the electron spins.
This field is given by the sum of the external and the Overhauser field, which allows us to extract the nuclear spin
polarisation.  The asymmetry of the recorded sets of spectra with respect to zero magnetic field is due to
nuclear spin-lattice relaxation. We have  taken it into account when fitting equation (1) to the data (black dashed lines in Fig. \ref{fig:sn}(b-e)).
For both samples and both signs of the nuclear spin temperature, the value of the
local field was found to be $B_L=12\pm 2$~G.
%
%%%This field is given by the sum of the external and the Overhauzer field, so that:
%%%\begin{equation}
%%%\nu=\gamma_e(B+B_N)=\gamma_e(B+b_NP_N),
%%%\label{nu}
%%%\end{equation}
%%%where $\gamma_e=0.64$~MHz/G is the gyromagnetic ratio of the electrons in the conduction band of GaAs\cite{Ryzhov2015}.
%%%%
%%%Thus, by measuring $\nu$ we obtain the value of the nuclear spin polarisation.
%%%%
%%%Because each field scan takes $100$~s ($10$ times longer than in the case of the Faraday rotation measurements), the spin-lattice relaxation of the NSS is not negligible on this time-scale.
%%%%
%%%It  manifests itself in the asymmetry of the recorded sets of spectra with respect to zero magnetic field.
%%%%
%%%For the quantitative comparison with the predictions by Eqs. 1,\ref{nu} we measured the magnetic field-dependent relaxation times
%%%in an independent set of experiments \cite{arXiv}.
%%%%
%%%The resulting fit of the data
%The result  is shown in the color maps by the black dashed lines (Fig. 3b-e).
%
% that is
%far above the well-known value $B_{dd}=1.5$~G given by the dipole-dipole coupling within NSS in GaAs \cite{Paget77}.
%
 Thus, the NSS does obey the prediction of the thermodynamic theory expressed by Eq. (1), but  value of the local field is surprisingly large, $B_L\approx 10$~G.
Indeed,  the spin-spin interactions in GaAs are dominated by magnetic dipole-dipole coupling, which yields
a much weaker local field $B_{dd}=1.5$~G\cite{Paget77}.

To elucidate the origin of this striking discrepancy, we performed spin noise measurements with the bulk GaAs layer without  a microcavity, Sample C (Fig. \ref{fig:sn}(f-g)).
Although the signal is much weaker,
%and the spectral line is much broader, presumably due to the inhomogeneity of the nuclear field across $20$~$\mu$m-thick layer, the field dependence of the spin noise resonance is well resolved.
%
the best fit using Eqs.~(1) and taking into account  spin-lattice relaxation during the measurement
yields $B_L=2$~G and $\Theta_{N0}=\pm 4$~$\mu$K \footnote{see details in Supplemental Material.}.
%%
%To further reveal the differences  in the NSS behaviour in bulk GaAs and in the GaAs-based microcavities,
%we show in Figs. 2c-e and 3c-h the results of the best fit assuming $B_L=2$~G.
%
%
%The strength of the local field is an important parameter of the NSS, because it determines the efficiency of the spin cooling via
%adiabatic demagnetisation process.
%%
%As follows from Eq. 1, the lowest spin temperature that can be achieved by adiabatic demagnetisation to zero field from
%$\Theta_{Ni}$ at $B_i$
%is given by $\Theta_{N0}=\Theta_{Ni} B_L/B_i$.
%%
%Therefore, after an equivalent pumping,  lower spin temperatures should be achieved in bulk sample, than in
%the microcavities.
%%
%Fig. 4 presents the experimental data with
%lowest zero-field spin temperatures obtained after $150$~minutes of pumping in Sample A
%and  $15$ minutes of pumping in Sample C under magnetic field $B_z^p=150$~G.
%%
%The peak in the spin noise spectra measured in the presence of the slowly varying magnetic field
%follows  Eq. 1 with $\Theta_{N0}=-2$~$\mu$K   for Sample C ($B_L=2$~G) and
%  $\Theta_{N0}=6$~$\mu$K for Sample A ($B_L=12$~G), respectively.
%%
%We have checked that the efficiency of the optical cooling followed by demagnetisation to zero field does not depend on the sign of the spin temperature.
%%
%These values of $\Theta_{N0}$ are similar to the record spin temperatures obtained in bulk GaAs \cite{Kalevich82}, but are well above the record-low nuclear spin temperatures achieved
% in metals
%$\Theta_{N0}\sim 100$~pK \cite{Oja}.
%%
%
This comparison shows unambiguously the  enhanced value of local field in the microcavities, compared to that in the bulk GaAs.
%
%establishment of the  spin temperature in
%the GaAs layers embedded in a microcavity, despite the local field well above the well-known effective field induced by the spin-spin  interactions between nuclei.
%
%
%The energy of a nuclear spin in $B_L$ is given by the product of a pair of the spin components
%$I_i I_j$ that can belong to either the same or two different nuclei;
%
Within the thermodynamic description of the NSS, the local field which enters  Eqs. (1)  is defined as \cite {Goldman}:
\begin{equation}
B_L^2=Tr(H_{S}^2)/Tr(M_{B}^2),
\label{BL}
\end{equation}
where $H_{S}$ is the Hamiltonian of all nuclear spin interactions, excluding Zeeman part (typically it includes the magnetic dipole-dipole interactions, and the indirect exchange),
and $M_B$ is the parallel to the magnetic field component of the nuclear magnetic moment.
%
%Spin-spin interactions usually include magnetic dipole interactions, and indirect exchange.
%
 In n-GaAs, magnetic dipole-dipole interaction is well-studied, and $B_L=2$~G measured in bulk GaAs agrees well with the previous estimations for $B_{dd} $ \cite{Paget77}.
 %
% It is unlikely that these interactions are different in microcavities.
 %
% In addition,
% The  value of the local field $B_L>>B_{dd}$ measured in the microcavities does not depend on the donor concentration.
 %
%This  excludes any interactions involving charge carriers from the list of possible contributions to the local field.
%

%The local field determines  the capacity of the NSS to store the energy in the internal degrees of freedom.
%%
%In the same manner, Zeeman energy is stored in the magnetisation that builds up in the presence of the external magnetic field.
%
%The component of the local field resulting from the dipole-dipole  interactions is responsible  not only for  the  energy storage, but also for the mixing between Zeeman (magnetic) energy reservoir and the internal energy reservoir.
%%
%Thus, it s a key quantity that controls the establishment of the thermodynamic equilibrium within the NSS.
%

The only plausible explanation for the unexpectedly strong local field detected in microcavities is the quadrupole splitting $h\nu_Q$ of the nuclear spin states induced by an uniaxial strain.
In Eq.(\ref{BL})  it can be accounted for   by introducing $H_S=H_{dd}+H_Q$,
where $H_Q$ is the Hamiltonian of the quadrupole interaction
\begin{equation}
H_Q=\sum_{i=1}^{3}{\frac{h\nu_{Q}^i}{2}(\hat{I}_z^2-\frac{I(I+1)}{3})}.
\label{HQ}
\end{equation}
Here the index $i$ stands for the summation over the three isotopes  ($^{69}Ga$,  $^{71}Ga$,
$^{75}As$),
and $\hat{I_z}$ is the  projection on the nuclear spin operator on the growth (strain) axis.
Using equation (\ref{BL}) and the parameters of strain-induced quadrupole splittings in GaAs \cite{Petrov}, one can estimate that the strain as weak as $0.01\%$ induces the local field
$B_L=10$~G in GaAs  \cite{Eickhoff2003}.
% \footnote{Similar quadrupole splittings were measured in multi-quantum well structures in Ref. \onlinecite{Eickhoff2003}}.
%.

Because
%this quadrupole-induced local field
$B_L>>B_{dd}$, it is the quadrupole interaction that determines the capacity of the NSS to store the energy in the internal degrees of freedom.
 %%
%In the same manner, Zeeman energy is stored in the magnetisation that builds up in the presence of the external magnetic field.
%%
%This means that the heat capacity of the NSS  is $50$ times larger in the microcavity samples than in bulk GaAs,
%because $C_N\propto B_L ^2$\cite{Goldman} .
%
But in contrast with dipole-dipole interaction, the quadrupole interaction  does not provide any coupling between the spins, and can not establish the thermodynamic equilibrium within the NSS.
Indeed, in quantum dots, where  strong quadrupole-induced local fields have been reported, nuclear spin temperature
failed to establish \cite{MaletinskyNatPhys2009}.
%
%The main manifestations of this effect were the remanent magnetisation at zero magnetic field,
%and the irreversibility of the magnetisation in the course of consecutive scans of the magnetic field across zero.
%%
%This is in striking contrast with our results.
% that show neither any loss of magnetisation, nor remanent magnetisation in zero magnetic field.
%
From our data we can estimate the lower limit of $50$~G for the {\it mixing field} $B_m$,  at which Zeeman
%(for each of three isotopes)
and internal energy reservoirs come to equilibrium between each other, so that the NSS can be described by the unique spin temperature \cite{Oja} (see Supplemental material).
%We measured  nuclear magnetisation   in Faraday rotation (Fig. 2) and
% in  the spin noise experiments  (Fig.6a) while scanning magnetic field between $50$ and $-50$~G.
%
  %Our data  show neither any loss of magnetisation, nor remanent magnetisation in zero magnetic field.
 %
\begin{figure*}
 \begin{center} \resizebox{6.5in}{!}{ \includegraphics{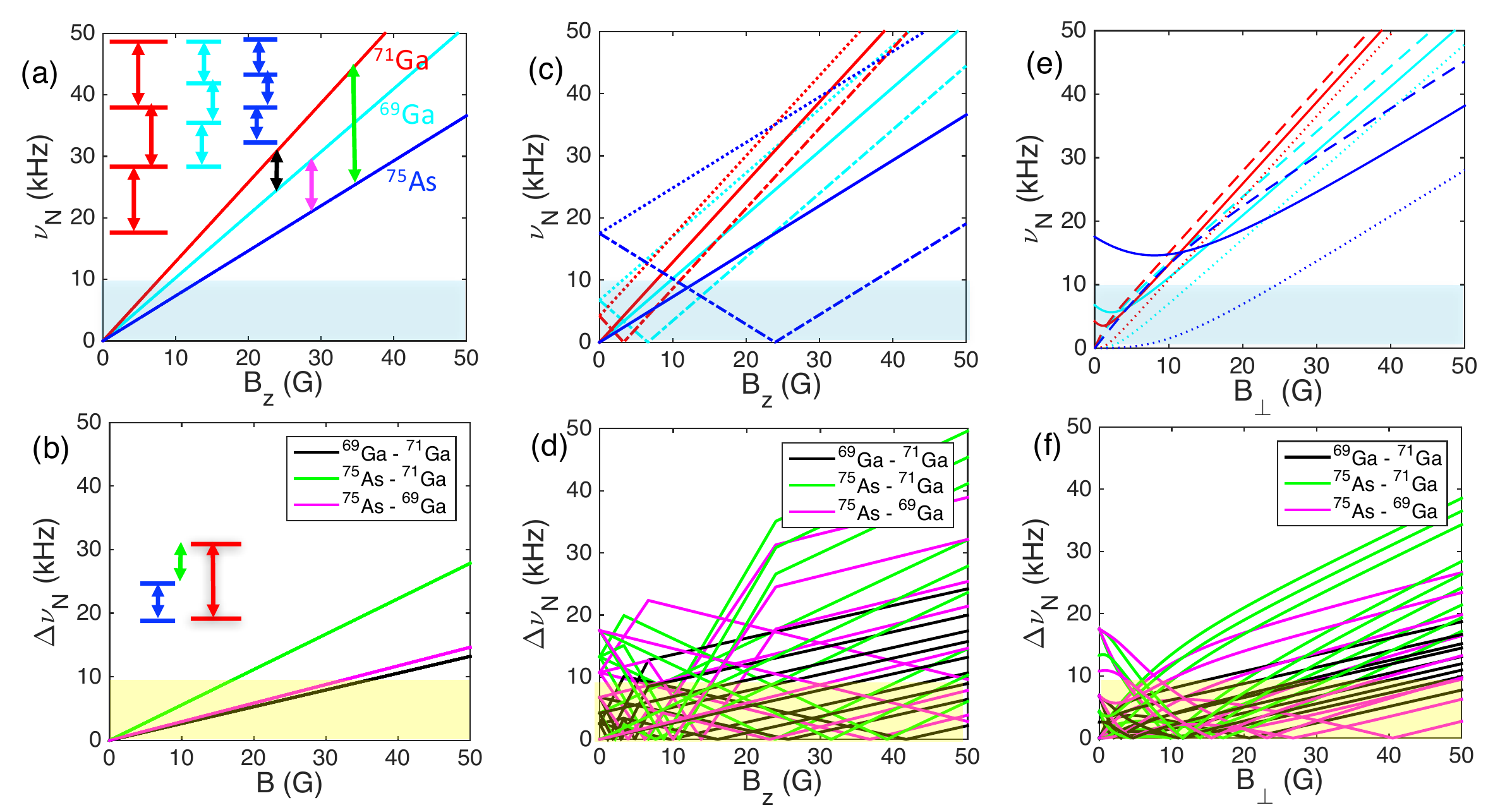} }
  \end{center}
  \caption{ (a) Nuclear spin flip transition frequencies $\nu_N$ for
 three GaAs isotopes, and  (b) the differences $\Delta\nu_N$ between them  as functions of magnetic field in the absence of the quadrupole splitting.
  (c, d) Same as  (a) and  (b), respectively, but in the presence of the quadrupole spitting in $z$-direction.
   (d, f), Same as   (c) and  (d), respectively, but the magnetic field is applied in the plane of the structure.
% Two inequivalent directions of the magnetic field are shown.
The blue area $\nu_N<8$~KHz in (a, c, d)  shows the extent of splittings in the local magneto-dipole field and indicates the range of external fields where mixing is possible within each individual isotope.
Similarly, the yellow area in (b, d, f) indicates the range of magnetic fields where mixing becomes possible if assisted by inter-isotope transitions.
}
\label{fig:forest}
 \end{figure*}
%
%
% The observed adiabaticity of the nuclear remagnetisation may have either thermodynamic or
% quantum mechanical (Ehrenfest) interpretation \cite{AbragamProctor}. However, in Faraday configuration Zeeman and quadrupole Hamiltonians commute, which makes impossible any energy transfer without the involvement of the dipole-dipole interaction. The later has a stochastic nature. This justifies the interpretation of our experiments in thermodynamic terms.

The question remains, how can the thermodynamic equilibrium be established under magnetic field $B_m>50$~G, much larger
than the characteristic field of the dipole-dipole interaction $B_{dd}=1.5$~G?
We suggest that this is made possible by the multi-isotope nature of the  NSS in GaAs.
The difference in the quadrupole splittings and gyromagnetic ratios between the three isotopes yields  a
rich variety of possible inter-isotope
 flip-flop transitions.
These transitions frequencies $\nu_N$ are illustrated in Fig.  \ref{fig:forest} as functions of the magnetic field in the absence (Fig.  \ref{fig:forest}a) and in the presence of the quadrupole splitting of the nuclear spin states along $z$-axis (Fig.  \ref{fig:forest} (c, e)).
The spin flip-flop transitions involving  different isotopes ensure the
energy transfer between the Zeeman and
quadrupole energy reservoirs, with total energy conservation of the NSS.
These  transitions are  broadened by dipole-dipole interactions.
% $\delta \nu=8$~kHz.
%
It is usually assumed \cite{Oja} that the efficient equilibration of energy reservoirs is ensured
at  detuning  from the resonance less than $\delta \nu_N=5 B_{dd}/(2\pi \langle \gamma_N \rangle=8$~kHz.
One can see in Fig.~\ref{fig:sn}d,f, that for both orientations of the magnetic field, the transitions involving such a small detuning are available  at $B<50$~G, and the mixing remains as efficient as in the absence
of the quadrupole splitting  (Fig. \ref{fig:sn}(b)).
%The broadening $\delta \nu$ is indicated at several values of magnetic field.%Nevertheless, the ratio $\Theta_{N0}/\Theta_L=2\cdot10^{-6}$, the
%
%They ensure the thermalisation of the NSS, that is the establishment of the common nuclear spin temperature
%for Zeeman, quadrupole, and internal energy reservoirs.

 %We have demonstrated a non-perturbative optical tool for the nuclear spin calorimetry.
 %
 Our results show that the strain-induced nuclear quadruple splittings in semiconductor microcavity do not hinder the establishment of the thermodynamic equilibrium within the nuclear spin system.
 The quadrupole effects result in the increase of the local field, indicating that the heat capacity of the
 NSS is dominated by the quadrupole energy reservoir.
The energy transfer between the Zeeman and quadrupole reservoirs during adiabatic demagnetisation is made possible by dipole-dipole interaction via spin flip-flop transitions involving  different isotopes.
 Thus, deep cooling of the NSS down to microKelvin temperature range via adiabatic demagnetisation is possible in photonic microstructures.
This paves the way towards realisation of nuclear magnetically ordered states  and their applications, including spin-photon interfaces with reduced thermal noise.
\begin{acknowledgments} This work was supported
 by the joint grant  of the Russian Foundation for Basic Research (RFBR, Grant No. 16-52-150008)   and
 National Center for Scientific Research (CNRS, PRC SPINCOOL No. 148362), as well as French National Research Agency (Grant  OBELIX, No. ANR-15-CE30-0020-02). IIR, VSZ and GGK  acknowledge Russian Foundation for Basic Research (grant No. 17-12-01124) for the financial support of their experimental work.
 \end{acknowledgments} 
 
 \pagebreak \widetext
\begin{center}
\textbf{\large Supplemental Material}
\end{center}
%%%%%%%%%% Merge with supplemental materials %%%%%%%%%%
%%%%%%%%%% Prefix a "S" to all equations, figures, tables and reset the counter %%%%%%%%%%
\setcounter{equation}{0} \setcounter{figure}{0}
\renewcommand{\theequation}{S\arabic{equation}}
\renewcommand{\thefigure}{S\arabic{figure}}
\section{Samples}
The studied microcavity structures consist of Si-doped GaAs $3\lambda/2$-cavity
 with electron concentrations $n_e=2\times 10^{15}$~cm$^{-3}$
 (Sample A) and $n_e=4\times 10^{16}$~cm$^{-3}$ (Sample B).
The front (back) mirrors are distributed Bragg reflectors composed
of $25$ ($30$) pairs of AlAs/Al$_{0.1}$Ga$_{0.9}$As layers, grown on
a $400~\mu$m thick GaAs substrate.
Due to multiple round trips in the cavity, the Faraday rotation (FR) is amplified  by a
factor of $N\sim 1000$, corresponding to the interaction length
$L=0.7$~mm (quality factor $Q\sim20000$ was measured by interferometric techniques)
Sample C is the bulk  $20$~$\mu$m-thick  GaAs layer   grown by liquid-phase epitaxy, with Si donor concentration of $n_d=4\times10^{15}$ cm$^{3}$.
All these samples have been studied previously \cite{Giri2013,Ryzhov2015,Kotur2016,Vladimirova2017}. %%
\section{Experimental techniques}
%Both are sensitive to the Overhauser field, an effective field created by the polarised nuclei and acting on electrons:
Both the spin-noise (SN) and FR techniques have been used previously for studies of the NSS \cite{Giri2013,Ryzhov2015,Vladimirova2017} and are described in in detail in Ref. \onlinecite{Vladimirova2017}.
They have an advantage of being virtually non-perturbative for the NSS, because pumping and measurement stages are separated in time, and
cooled NSS is  optically probed via the polarisation rotation of the light beam with photon energy
tuned below  (here $20$~meV) the band gap of the studied GaAs layer.
In a typical measurement, the sample is placed in a cold finger cryostat at $T=5$~K, $B_z=180$~G.
At the first stage,   it is
optically pumped during $3$-$15$ minutes by the circularly polarised laser diode with photon energy $1.57$~eV and power $P=10$~mW, focused on $1$~mm spot on the sample surface.
In the case of SN experiments, the transverse field $B_\perp$ is also applied during pumping.
After the pumping stage, we wait for $t_w=1$ minute before lowering down $B_z$ (Fig. 2(a), 3(a)), to be sure that  nuclear spins situated under the orbits of the donors and characterised by the relatively short $T_1$ do not contribute to the signal \cite{Giri2013,Ryzhov2015,Ryzhov2016}.
At the last preparation step, $B_z$ is lowering down to the value from which the measurement stage starts
($B_z=B_z^i=50$~G for FR and $B_z=0$ for SN).
FR and SN experiments mainly differ by the measurement stage.
In FR experiment, the rotation of the linearly polarised probe beam is detected in the presence of the slowly varying longitudinal magnetic field $B_z$.
In the SN experiment, the spin noise of the resident electrons is measured in the presence of the slowly varying transverse  magnetic field $B_\perp$ via the fluctuation spectrum of the Faraday rotation angle.
The probe beam has the photon energy $20$~meV below GaAs band gap, power of $0.5$~mW, and  is focused on $30$~$\mu$m spot on the sample surface.
%
%The advantage of the Faraday rotation is its sensitivity to the sign of the nuclear magnetisation, while the spin noise technique offers the access of the absolute value of the nuclear spin polarisation and temperature.
%
%

\subsection
{Faraday rotation} To extract nuclear spin temperature and the local field from the Faraday rotation angle measured as a function of the slowly varying magnetic field $B_z$ (the duration of each scan is  $\sim 10$~s so that $dB/dt=10$~G/s), we proceed as follows.
First, we subtract the external field contribution from the total signal. This contribution to the signal remains unchanged
for all consecutive scans and depends linearly on the magnetic field.
The remaining part of the FR is induced by the Overhauser field $B_N$,
which is proportional to the nuclear spin polarisation $P_N$.
 \begin{equation}
\phi_N=B_N V_N L=b_N P_N V_N L,
\label{phiN1}
 \end{equation}
where $b_N=5.3$~T is the Overhauser field produced by the fully polarised nuclear spins \cite{OpticalOrientation}, $V_N$ is the nuclear Verdet constant, $L$ is the effective optical length of the sample accounting for
by multiple round trips of  light in the cavity \cite{Giri2013}.
Therefore,  from Eqs. (1) we get:
 \begin{equation}
 \phi_N = \phi_{Ni}B/\sqrt{B^2+B_L ^2}
 \label{phiN}
 \end{equation}
 where $ \phi_{Ni}$ is the Faraday rotation angle at the saturation field $B_z^i$.
 Fitting $\phi_N$  to equation (\ref{phiN}) we determine $B_L=8\pm2$~G.
Using the values  of $V_N=0.1$~mrad/G/cm  and $L=0.7$~mm
 determined in our previous work \cite{Giri2013} we also extract  $\Theta_{N0}=-6$~$\mu$K in sample B (Fig. 2 (e) in the main text)
 from Eqs. (\ref{phiN1}), (\ref{phiN}) and (1).
 The values of $B_L$ and $\Theta_{N0}$ extracted from FR measurement are averaged over the
 crystal volume, since the signal is given by the electron band spin splitting \cite{Giri2013}.
 %
 %the
%knowledge of the nuclear Verdet constant $V_N$ and the effective optical length of the sample $L$.
%
%In our previous work, we  have estimated $V_N=0.1$~mrad/G/cm  and $L=0.7$~mm in Sample B.

\subsection {Spin noise spectroscopy}  The electron spin noise spectrum exhibits a
pronounced peak at the electron Larmor  frequency  $\nu$ corresponding to the total ($B_{\perp}$ and $B_N$) field , so that: \cite{Ryzhov2015,Ryzhov2016}
%
%The transverse component of the field is tuned across zero at the rate of $\sim 0.5$~G/s.
%
%The narrow peak in the spectra appears at the frequency $\nu$ of the electron Larmor precession in the
%total effective magnetic field acting upon the electron spins.
%
%This field is given by the sum of the external and the Overhauzer field, so that:
\begin{equation}
\nu=\gamma_e(B_{\perp}+B_N)=\gamma_e(B_{\perp}+b_NP_N),
\label{nu}
\end{equation}
where $\gamma_e=0.64$~MHz/G is the gyromagnetic ratio of the electrons in the conduction band of GaAs\cite{Ryzhov2015}.
Thus, by measuring $\nu$ as a function of $B_{\perp}$ and fitting equations (1) and (\ref{nu}) to the data we obtain the values of $\Theta_{N0}$ and $B_L$.
Because each field scan takes $100$~s ($10$ times longer than in the case of the Faraday rotation measurements), the spin-lattice relaxation of the NSS is not negligible on this time-scale.
It  manifests itself in the asymmetry of the recorded sets of spectra with respect to zero magnetic field.
For the quantitative comparison with the theory predictions given by Eqs. (1) and (\ref{nu}) we measured the magnetic field-dependent relaxation times
in an independent set of experiments \cite{Vladimirova2017}.
Note that in metallic samples the SN signal is mediated by the electron gas, and is therefore
contributed by all the nuclei. In the insulating samples, only the nuclei situated under the
donor orbits can be detected. However, the polarisation of the nuclear spins situated in the core of the donor orbit decays rapidly, and vanishes during $t_w$.
Thus the SN signal comes from the nuclei situated in the periphery of the donor orbits,
so that the extracted  $\Theta_{N0}$ and $B_L$ are close to those of the bulk nuclei.
%Here is a description of a specific method used.  Note that the
%subsection heading ends with a full stop (period) and that the
%command is \verb|\subsection{}| not \verb|\subsection*{}|.
%
\section{Estimation of the mixing field}
The mixing field $B_m$, is the field at which Zeeman
(for each of three isotopes)
and internal energy reservoirs come to equilibrium between each other \cite{Oja}.
%
%Indeed, dynamic nuclear polarisation at strong field cools down only the Zeeman reservoirs.
Only the Zeeman reservoirs can be cooled down via dynamic nuclear polarisation at strong field.
By measuring nuclear polarisation, we get access to the average energy of the Zeeman reservoirs
$\langle E_Z\rangle=B P_N \langle \hbar \gamma_N \rangle$.
During adiabatic demagnetisation,  energy transfer and the thermalisation between Zeeman and internal energy
 reservoirs is achieved  at $B=B_m$.
At this field, a part of Zeeman energy is transferred to the internal energy reservoir, which results in the modification
of the nuclear polarisation.
The nonadiabaticity of this process would lead to a deviation from Eqs. (1), quantified by the nonadiabaticity factor
$f_{na}=B_m/\sqrt{B^2+B_m^2}$ \cite{Oja}.
Comparing the magnetisation measured at $B=50$~G before and after the passage through zero field,
we have not observed any difference within the experimental precision of $2\%$.
This yields $f_{na}>0.98$, and therefore $B_m>50$~G.

\bibliography{LocalField}
\end{document}